\documentclass[11pt,en]{elegantpaper}

\title{Constructing self-referential instances for the clique problem}
\usepackage{array}
\usepackage{CJKutf8}

\usepackage{amssymb, amsthm, amsmath}
\usepackage{subfiles}
\author
{Jiaqi Li \\
Northeast Normal University \\
lijq515@nenu.edu.cn\\
\and Shuli Hu \\
Northeast Normal University \\
husl903@nenu.edu.cn\\
\and Xianxian Li \\
Guangxi Normal University \\
lixx@gxnu.edu.cn\\
\and Minghao Yin \thanks{Corresponding author}\\
Northeast Normal University\\
ymh@nenu.edu.cn
}

\date{}

\begin{document}
\begin{CJK}{UTF8}{gbsn}
\maketitle

\begin{abstract}
In this paper, we propose constructing self-referential instances to reveal the inherent algorithmic hardness of the clique problem. First, we prove the existence of a phase transition phenomenon for the clique problem in the Erd\H{o}s--R\'enyi random graph model and derive an exact location for the transition point. Subsequently, at the transition point, we construct a family of graphs. In this family, each graph shares the same number of vertices, number of edges, and degree sequence, yet both instances containing a $k$-clique and instances without any $k$-clique are included. These
two states can be transformed into each other through a symmetric transformation that preserves the degree of every vertex. This property explains why exhaustive search is required in the critical region: an algorithm must search nearly the entire solution space to determine the existence of a solution; otherwise, a counterinstance can be constructed from the original instance using the symmetric transformation. Finally, this paper elaborates on the intrinsic reason for this phenomenon from the independence of the solution space.
\end{abstract}

\bigskip

\section{Introduction}

Over the past decade, stronger conjectures than P $\neq$ NP—such as the Strong Exponential Time Hypothesis—have attracted increasing attention (e.g.,~\cite{vassilevska2015hardness,cygan2016problems}). Recently, Xu and Zhou~\cite{xu2025sat} drew inspiration from Gödel’s ideas and demonstrated that brute-force computation is indeed unavoidable in certain cases by constructing self-referential instances. Gödel was inspired by the liar paradox to construct the famous self-referential proposition. The paradox itself is self-contradictory and thus not permitted in mathematics. The essence of a self-referential object lies in its inclusion of two contradictory aspects without being self-contradictory. Specifically, the liar paradox is self-contradictory because the proposition $F$ cannot be equivalent to $\neg F$. In contrast, Gödel’s self-referential proposition is not self-contradictory, as $F$ can be equivalent to the unprovability of $F$. In a certain sense, Gödel constructed an extremely difficult mathematical proposition, because he used the diagonalization method to prove that this proposition is forever unprovable within any finite formal system. The core purpose of constructing self-referential objects is to determine clear boundaries for reasoning. The essence of algorithm design is to create reasoning rules based on the characteristics of a problem, so constructing extremely difficult self-referential instances is a viable method for determining the limits of algorithms.

To gain a better understanding of self-referential objects, we consider an intuitive example: a coin inherently has both a heads and a tails side. We can say that heads and tails form a contradictory pair, since at any given moment the coin must be either heads-up or tails-up. On the other hand, the coin itself is not self-contradictory, because it cannot be both heads-up and tails-up simultaneously. These two states can be transformed into each other through a single action—flipping the coin. This property of containing two contradictory aspects and allowing transformation between them through a single action prevents rule-based reasoning; instead, one must inspect the object based on its semantic definition to determine its state. If the coin is placed inside a box, its nature dictates that to determine its state, one must open the box and check. Of course, this conclusion also relies on certain assumptions, such as the box being sealed and opaque. Clearly, the conclusion reflects the nature of the coin itself rather than the assumptions. For instance, if the box contains water and we want to determine whether it is in a liquid or solid state, even if the box is sealed and opaque, we can infer the state based on temperature without opening the box. This is because the transformation of water into ice and ice into water are two opposite processes—one involves cooling, while the other involves heating. Imagine if there were a single process that could both freeze water and melt ice, then reasoning would become impossible, and we would have no choice but to open the box and check directly.

Now we consider another simple example: a set composed of positive and negative numbers, and the task is to determine whether a number in the set is positive or negative. Obviously, this is easy to do based on the semantic definitions of positive and negative numbers. The question now is whether this task can still be accomplished without relying on those semantic definitions. In some cases, it is possible. For instance, if all the positive numbers in the set have absolute values greater than 2, while all the negative numbers have absolute values less than 2, then we can distinguish between them using absolute value alone, without referring to the semantic definitions of positive and negative. Clearly, positive and negative numbers can be converted into each other by multiplying by -1. We define multiplying a set by -1 as multiplying each element in the set by -1. If the positive and negative numbers in a set are symmetric—for example, \{1, -1, 3, -3, 4, -4\}—then multiplying the set by -1 results in the same set. In such cases, we can only distinguish the numbers based on the semantic definitions of positive and negative.

From the above analysis and discussion, it is not hard to see that the computational hardness of self-referential instances stems from the inherent characteristics of the instances themselves. For a given problem, if self-referential instances of that problem can be constructed, then the diagonalization method can be used to prove that, in the worst case, brute-force computation is required. In a certain sense, constructing self-referential instances is both the key and the difficulty of the proof. Once such instances are established, applying the diagonalization method for contradiction is not difficult. For decades, prior to the work of Xu and Zhou~\cite{xu2025sat}, no one had applied Gödel’s method to proving lower bounds in computational complexity. One crucial reason may have been the lack of understanding of how to construct self-referential instances. Professor Wigderson once said~\cite{wigderson2019mathematics}: ``\textit{Concluding, I view the mystery of the difficulty of proving (even the slightest non-trivial) computational difficulty of natural problems to be one of the greatest mysteries of contemporary mathematics}". Computation is also reasoning. Proving that a problem is computationally hard is equivalent to proving that reasoning about the problem is hard—and proof itself is a form of reasoning. Therefore, if one manages to prove that reasoning is hard, it implies that reasoning is not so hard after all, creating a contradiction. To break this self-referential loop, constructing self-referential instances is a feasible (and perhaps the only) approach. It should be noted that Professor Mulmuley~\cite{mulmuley2011p} has expressed a similar view on the inherent conceptual difficulty of proving P vs. NP in his geometric complexity theory (GCT). Additionally, Professors Fellows and Rosamonda~\cite{P_Not_NP} have put forward a novel conjecture, the Hard Puzzles Conjecture (HPC), which aims to investigate computational complexity through the design of intrinsically hard instances. Inspired by the work of Xu and Zhou~\cite{xu2025sat}, this paper constructs self-referential instances for the well-known clique problem.

\section{Preliminaries}


In 1959, Erdős and Rényi introduced the random graph model $G(n, m)$\cite{erd6s1960evolution}, where $n$ represents the number of vertices and $m$ represents the number of edges. Their work demonstrated that as $n \to \infty$, increasing $m$ causes the probability of the graph possessing specific properties (such as connectivity) to transition sharply from $0$ to $1$. In 1978, Bollobás and Thomason further established that random graphs exhibit monotonic threshold functions for graph properties. This means there exists a critical interval where the probability that the property holds transitions rapidly from 0 to 1, and this interval converges to a single point as $n$ increases. If a graph $H$ contains a $k$-clique, then any graph $G$ containing $H$ as a subgraph also possesses a $k$-clique, these two properties—according to the theory established by Ehud Friedgut \cite{friedgut1999sharp}—imply that the existence of a complete subgraph constitutes a graph property. Consequently, the property of whether a graph contains a $k$-clique also exhibits a phase transition phenomenon.

Furthermore, the stringent definition of a clique as a complete subgraph has motivated the introduction of relaxed structures such as the $k$-plex\cite{seidman1978graph}. Formally, in an undirected graph $G=(V,E)$, a $k$-plex is a subgraph $S$ where each vertex is non-adjacent to at most $k$ other vertices within $S$. 


Let $\mathbb{G}$ denote the set of undirected graphs with $|V|=n$ vertices and $|E|=m$ edges. Let $Q = \frac{1}{2}{n(n-1)}$. If $\forall G \in \mathbb{G}$, the probability that a uniformly selected graph from $\mathbb{G}$ is exactly $G$ equal to $\frac{1}{\binom{m}{Q} }$, then $\mathbb{G}$ is termed as a uniform undirected random graph set with $n$ vertices and $m$ edges, and $G$ is called a uniform random undirected graph. 

A vertex subset $T \subseteq V$ is called a clique of an undirected graph $G(V, E)$ if and only if every pair of distinct vertices $u, v \in T$ satisfies $(u, v) \in E$. A clique in $G(V, E)$ that contains exactly $k$ vertices is termed a $k$-clique. 

Define the edge density as $\rho = \frac{2m}{n(n-1)}$, which quantifies the sparsity of the graph. If there exists a function $f(p)$ such that, in the asymptotic regime $n \to \infty$, the following holds:

\begin{equation}
  \lim\limits_{n \to \infty}\mathbb{P}(A)=
  \begin{cases}
    0 , \text{If} \; \rho < W(n)\\
    1 , \text{If} \; \rho \geq W(n) \\
  \end{cases},
\label{P_A}
\end{equation}
where $A$ denotes an event that the graph $G(V, E)$ contains a $k$-clique and $W(n)$ is termed the critical threshold for $k$-clique. 

In phase transition analysis, the introduction of integer-valued random variables to enumerate configurations with specific properties in a random graph $G(V, E)$ and compute their expectations can simplify the proofs by bypassing redundant probabilistic arguments. 
Let $X$ be a non-negative integer-valued random variable such that $X=0$ if the random graph $G(V, E)$ doesn't contains $k$-clique. Formally, we introduce the first-moment method. 

\begin{lemma}
\label{First-moment}
For any non-negative integer-valued $X$, if $\lim\limits_{n \to \infty}\mathbb{E}[X]=0$, it holds that
  \begin{equation}
    \lim\limits_{n \to \infty} \mathbb{P}[X \geq 1]=0.
  \end{equation}
\end{lemma}

\begin{proof}
Since $X$ is a non-negative random variable, we could apply Markov’s inequality:
  \begin{equation*}
    \mathbb{P}[X \geq C] \leq \frac{\mathbb{E}[X]}{C}.
  \end{equation*}
Choosing $C = 1$, when $n \to \infty$, we have:
\begin{equation*}
   \lim\limits_{n \to \infty}\mathbb{P}[X \geq 1] \leq \lim\limits_{n \to \infty}\mathbb{E}[X] = 0.
\end{equation*}
\end{proof}

Then, $\mathbb{E}[X] \to 0$ implies the absence of $k$-clique in $G(V, E)$. 

Lemma \ref{First-moment} rigorously establishes that $G(V, E)$ asymptotically almost surely contains no $k$-clique. However, an analogy to the first-moment method to prove $\lim\limits_{n \to \infty}\mathbb{E}[X]= \infty$ cannot establish $\mathbb{P}[X=0]=0$, i.e., it fails to rigorously demonstrate the certain existence of a $k$-clique in the graph. A diverging expectation $\mathbb{E}[X] \to \infty$ doesn't guarantee the presence of $k$-clique. For example, suppose $\mathbb{P}[X=0]=0. 5$ and $\mathbb{P}[X=n]=0. 5$. Then $\mathbb{E}[X] = 0.5n \to \infty$ when $n \to \infty$, yet $\mathbb{P}[X=0]=0. 5$, which does not vanish. To address this, we introduce the second-moment method. 

\begin{lemma}
\label{Second-moment}
If the variance of $X$ satisfies $\lim\limits_{n \to \infty}\mathbb{E}[X]= \infty$ and 
\begin{equation}
  Var[X]=o(\mathbb{E}^2[X]),
\end{equation}
it holds that
\begin{equation}
  \lim\limits_{n \to \infty}\mathbb{P}[X=0] \leq 0.
\end{equation}
\end{lemma}

\begin{proof}

For a non-negative discrete random variable $X$, if $\mathbb{E}[X] > 0$, we could apply Chebyshev's inequality with $\epsilon = \mathbb{E}[X]$:
\begin{equation*}
  \begin{aligned}
   \frac{Var[X]}{\mathbb{E}^2[X]} &\geq \mathbb{P}[|X-\mathbb{E}[X]| \geq \mathbb{E}[X]] \\
   &= \mathbb{P}[X \leq 0] + \mathbb{P}[X \geq 2\mathbb{E}[X]] \\
   &\geq \mathbb{P}[X \leq 0] \\ 
   &\geq \mathbb{P}[X = 0].
  \end{aligned}
\end{equation*}
Since $Var[X]=o(\mathbb{E}^2[X])$, it holds that:
\begin{equation*}
  \begin{aligned}
    \lim\limits_{n \to \infty}\mathbb{P}[X=0] \leq \lim\limits_{n \to \infty}\frac{Var[X]}{\mathbb{E}^2[X]} = 0.
  \end{aligned}
\end{equation*}
\end{proof}


In this article, we assume that $k = O(\ln (n))$. 

\section{Analysis of Critical Thresholds}

Let $\Omega$ denote the set of all possible $k$-vertex subsets in $G(V, E)$. For each $\alpha \in \Omega$, a binary random variable $X_{\alpha}$ such that $X_{\alpha}=1$ if the subgraph induced by $\alpha$ forms a clique, and $X_{\alpha}=0$ otherwise. 

Let $X=\sum_{\alpha \in \Omega}X_{\alpha}$ denote the number of $k$-clique in an undirected graph $G(V, E)$ with $|V|=n$. 

In the random graph $G(V, E)$, the expectation of $X$ is given by
\begin{equation}
    \mathbb{E}[X]=\mathbb{E}[\binom{n}{k} \cdot  X_\alpha]=\binom{n}{k} \cdot  \mathbb{E}[X_\alpha],
\label{EX_and_X_a}
\end{equation}
where $\mathbb{E}[X_\alpha]$ represents the probability that a uniformly random subset of $k$ vertices forms a clique. According to Lemma \ref{First-moment}, if a critical threshold $W(n)$ exists, then for $m < W(n)$, we have $\mathbb{E}[X] \to 0$ as $n \to \infty$.

\begin{lemma}
\label{Leq_phase}
For a random graph $G(V, E)$, if $m < \frac{n(n-1)}{2} \cdot n ^{-\frac{2}{k-1}}$, then when $n \to \infty$, we have $E[X] \to 0$, implying the random graph asymptotically almost surely contains no $k$-clique. 
\end{lemma}

\begin{proof}

Let

\begin{equation*}
\beta = \frac{k(k-1)}{2}, g=\frac{n(n-1)}{2}.
\end{equation*}
By Stirling's approximation
\begin{equation*}
n! = \sqrt{2\pi n} \left(\frac{n}{e}\right)^n e^{\frac{\theta_n}{12n}}, \theta_n \in [0, 1],
\end{equation*}
applying Equation (\ref{EX_and_X_a}) derives that
\begin{equation*}
\footnotesize
\begin{aligned}
\mathbb{E}(X_\alpha) &= \binom{\beta}{\beta} \cdot \binom{g - \beta}{m - \beta} \cdot \frac{1}{\binom{g}{m}} \\
&= \frac{(g-\beta)!}{(m-\beta)!(g-m)!} \cdot \frac{m!(g-m)!}{g!}\\
&= \frac{(g-\beta)!m!}{g!(m-\beta)!} \\
&= \frac{ \sqrt{2\pi(g-\beta)} (\frac{g-\beta}{e})^{g-\beta} }{ \sqrt{2\pi g} (\frac{g}{e})^{g} } \cdot \exp \{{ \frac{\theta_{g-\beta}}{12(g-\beta)} - \frac{\theta_{g}}{12g} }\} \\
&\cdot \frac{ \sqrt{2\pi m}(\frac{m}{e})^{m} (\frac{m}{e})^{m} }{ \sqrt{2\pi (m-\beta)}(\frac{(m-\beta)}{e})^{(m-\beta)} } \cdot \exp \{{ \frac{\theta_{m}}{12m} - \frac{\theta_{m-\beta}}{12(m-\beta)} }\}\\
&\approx \sqrt{\frac{mg-\beta m}{mg-\beta g}} \cdot \frac{(g-\beta)^{g-\beta}m^m}{g^{g}(m-\beta)^{m-\beta}} \\
&= e^{-\beta + \beta} \left(\frac{m}{g}\right)^\beta \cdot \sqrt{1+\frac{\beta(g-m)}{(m-\beta)g}}\\
&= \left(\frac{m}{g}\right)^\beta \sqrt{1+\frac{\beta(g-m)}{(m-\beta)g}}.
\label{E_X_alpha}
\end{aligned}
\end{equation*}

Therefore, the expected number of $k$-clique is given by

\begin{equation}
\footnotesize
\begin{aligned}
\label{E_Xn_2}
\mathbb{E}[X] &= \binom{n}{k} \cdot \mathbb{E}(X_\alpha) \\
&\leq \frac{n^k}{k!} \cdot \mathbb{E}(X_\alpha) \\
&=\frac{1}{\sqrt{2\pi k}} \cdot \left(\frac{en}{k}\right)^k \cdot \sqrt{1+\frac{\beta(g-m)}{(m-\beta)g}} \cdot \left(\frac{m}{g}\right)^\beta \cdot \frac{1}{(1+o(\frac{1}{k}))}\\
&\leq\frac{1}{\sqrt{2\pi k}} \cdot \left(\frac{e}{k}\right)^k \cdot \sqrt{1+\frac{\beta(g-m)}{(m-\beta)g}} \cdot \left(n\left(\frac{m}{g}\right)^{\frac{k-1}{2}} \right)^k.\\
\end{aligned}
\end{equation}

Since the divergence rate of $\frac{1}{\sqrt{2\pi k}}\left(\frac{e}{k}\right)^k$ is asymptotically dominated by $n^k$, it suffices to analyze the term $\left(n\left(\frac{m}{g}\right)^{\frac{k-1}{2}}\right)^k$. Assuming $m=g\cdot n^{-\frac{2c}{k-1}}$ and defining $m_0(n)=g\cdot n^{-\frac{2}{k-1}}$, then the Eq. (\ref{E_Xn_2}) simplifies to:

\begin{equation}
\footnotesize
\begin{aligned}
\label{E_x_equ}
\mathbb{E}[X] &=\frac{1}{\sqrt{2\pi k}} \cdot \left(\frac{e}{k}\right)^k \cdot \sqrt{1+\frac{\beta(g-gn^{-\frac{2c}{k-1}})}{(gn^{-\frac{2c}{k-1}}-\beta)g}} \cdot \left(n\left(\frac{gn^{-\frac{2c}{k-1}}}{g}\right)^{\frac{k-1}{2}}\right)^k\\
&=o(k^{-k}) \cdot \sqrt{1+\frac{\beta g(1-n^{\frac{2c}{k-1}})}{(gn^{\frac{2c}{k-1}}-\beta)g}} \cdot [n^{1-c}]^k\\
&=o(k^{-k}) \cdot (n^{1-c})^k \cdot o(\frac{1}{n}).\\
\end{aligned}
\end{equation}

When $c > 1$, the above expression asymptotically tends to zero. Consequently, it can be rigorously concluded that $k$-clique are absent in the graph with high probability. 

\end{proof}

To prove the case when $m > n ^{-\frac{2}{k-1}}$, according to Lemma  \ref{Second-moment}, we need to introduce $\mathbb{E}^2[X]$.

\begin{lemma}
\label{Geq_phase}
For a random graph $G(V, E)$, if $m > \frac{n(n-1)}{2} \cdot n  ^{-\frac{2}{k-1}}$, then when $n \to \infty$, we have $\text{Var}[X] =  o(\mathbb{E}^2[X])$, implying the random graph asymptotically almost  surely contains a $k$-clique. 
\end{lemma}

\begin{proof}
By Lemma \ref{Second-moment}, we can computes $\mathbb{E}^2[X]$, which corresponds to enumerating all ordered pairs of $k$-clique and analyzing their co-occurrence probabilities. Let $w=2\beta-\frac{r(r-1)}{2}$, then
    \begin{equation*}
    \footnotesize
        \begin{aligned}
        \mathbb{E}[X^2]&=\mathbb{E}[X]+\sum_{\alpha \neq \beta \in \theta}\mathbb{E}[X_\alpha X_\beta] \\
        &=\mathbb{E}[X]+
        \sum_{r=0}^k \binom{n}{k}\cdot \binom{k}{r} \cdot \binom{n-k}{k-r} \cdot \binom{g - w}{m - w} \cdot \frac{1}{\binom{g}{m}}.\\
        \end{aligned}
    \end{equation*}
To establish the relationship between $Var(X)$ and $\mathbb{E}^2[X]$, we derive the following
\begin{equation}
\footnotesize
\label{E_x_2_equ}
    \begin{aligned}
    \frac{Var(X)}{\mathbb{E}^2[X]} &= \frac{\mathbb{E}[X^2] - \mathbb{E}^2[X]}{\mathbb{E}^2[X]} \\
    &= \frac{\mathbb{E}[X] + \binom{n}{k}
    \sum_{r=0}^k \binom{k}{r} \binom{n-k}{k-r} \binom{g - w}{m - w} \cdot \frac{1}{\binom{g}{m}}}{\mathbb{E}^2[X]} - 1\\
    &= \frac{1}{\mathbb{E}[X]} + \frac{\binom{n-k}{k} \binom{g-2\beta}{m-2\beta} \binom{g}{m} } {  \binom{n}{k} \binom{g-\beta}{m-\beta}^2 } \frac{k \binom{n-k}{k-1} \binom{g-2\beta}{m-2\beta} \binom{g}{m}}{\binom{n}{k} \binom{g-\beta}{m-\beta}^2}\\
    &+ \frac{ \sum_{r=2}^k \binom{k}{r} \binom{n-k}{k-r} \binom{g - w}{m - w} \binom{g}{m}}{\binom{n}{k} \binom{g-\beta}{m-\beta}^2 } - 1 \\
    &= \frac{1}{\mathbb{E}[X]} + \frac{ \binom{n - k}{k} + k\binom{n - k}{k - 1} }{ \binom{n}{k} } \; \frac{ \binom{g-2\beta}{m-2\beta} \binom{g}{m} }{ \binom{g-\beta}{m-\beta}^2 }\\
    &+ \frac{ \sum_{r=2}^k \binom{k}{r} \binom{n-k}{k-r} \binom{g - w}{m - w} \binom{g}{m}}{\binom{n}{k} \binom{g-\beta}{m-\beta}^2 } - 1 \\
    &\triangleq \frac{1}{\mathbb{E}[X]} + A_n \cdot  B_n + C_n - 1.\\
    \end{aligned}
\end{equation}

Here, we define
\begin{subequations}
\footnotesize
\begin{align}
A_n&=\frac{ \binom{n - k}{k} + k\binom{n - k}{k - 1} }{ \binom{n}{k} }, \label{A_n} \\ 
B_n&=\frac{ \binom{g-2\beta}{m-2\beta} \binom{g}{m} }{ \binom{g-\beta}{m-\beta}^2 }, \label{B_n}\\
C_n&=\frac{ \sum_{r=2}^k \binom{k}{r} \binom{n-k}{k-r} \binom{g - w}{m - w} \binom{g}{m}}{\binom{n}{k} \binom{g-\beta}{m-\beta}^2 }\label{C_n}.
\end{align}
\end{subequations}

For (\ref{A_n}), we have
\begin{equation*}
\footnotesize
\frac{\binom{n - k}{k}}{\binom{n}{k}}=\frac{\frac{(n-k)!}{k!(n-2k)!}}{\frac{n!}{k!(n-k)!}}=\Pi_{i=1}^k\frac{n-k-i+1}{n-i+1}.
\end{equation*}

By the Squeeze Theorem, observe that:
\begin{equation*}
\footnotesize
\left(1-\frac{k}{n-k+1}\right)^k=e^{-\frac{k^2}{n-k+1}} \to 1 \le \Pi_{i=1}^k\frac{n-k-i+1}{n-i+1} \le 1.
\end{equation*}
Thus, 
\begin{equation*}
\footnotesize
\lim_{n \to \infty} \frac{\binom{n - k}{k}}{\binom{n}{k}}=1.
\end{equation*}
For the second term , we have
\begin{equation}
\footnotesize
    \begin{aligned}
    \frac{k \binom{n - k}{k-1} }{\binom{n}{k}}&=k\cdot \frac{\frac{(n-k)!}{(k-1)!(n-2k+1)!}}{\frac{n!}{k!(n-k)!}} \\
    &=k^2\cdot \frac{(n-k)!(n-k)!}{n!(n-2k+1)!}\\
    &=\frac{k^2}{n-2k+1}\cdot \frac{(n-k)!(n-k)!}{n!(n-2k)!} \le \frac{k^2}{n+1}.
    \label{two_term}
    \end{aligned}
\end{equation}

Given $k =O(\ln (n))$, the Eq. (\ref{two_term}) vanishes asymptotically $\lim\limits_{n \to \infty} o(\frac{(\ln \;n)^2}{n}) \to 0. $ Combining both results, $\lim\limits_{n \to \infty}A_n=\frac{ \binom{n - k}{k} + k\binom{n - k}{k - 1} }{ \binom{n}{k} }=1+0=1$. 

For (\ref{B_n}), we have
\begin{equation*}
\footnotesize
    \begin{aligned}
    B_n&=\frac{ \binom{g-2\beta}{m-2\beta} \binom{g}{m} }{ \binom{g-\beta}{m-\beta}^2 } \\
    &=\frac{ \frac{(g-2\beta)!}{(m-2\beta)!(g-m)!} \frac{g!}{m!(g-m)!} }{ \frac{[(g-\beta)!]^2}{[(m-\beta)!]^2 [(g-m)!]^2 } } \\
    &=\frac{(g-2\beta)!(g)!(m-\beta)!(m-\beta)!}{(g-\beta)!(g-\beta)!m!(m-2\beta)!}\\
    &= \frac{(m-\beta)\sqrt{(g-2\beta)(g)}}{(g-\beta)\sqrt{m(m-2\beta)}}\\
    &\cdot \frac{(\frac{g-2\beta}{e})^{g-2\beta} (\frac{g}{e})^{g} (\frac{m-\beta}{e})^{2m-2\beta} }{ (\frac{g-\beta}{e})^{2sum-2\beta} (\frac{m}{e})^{m} (\frac{m-2\beta}{e})^{m-2\beta} }\\
    &= \sqrt{ (1-\frac{\beta}{g-\beta})(1+\frac{\beta}{g-\beta})} \cdot \sqrt{(1-\frac{\beta}{m})(1+\frac{\beta}{m-2\beta})}\\
    &\cdot \frac{(g-2\beta)^{g-2\beta}g^{g}}{(g-\beta)^{2sum-2\beta}} \frac{(m-\beta)^{2m-2\beta}}{m^m(m-2\beta)^{m-2\beta}},\\
    \end{aligned}
\end{equation*}
we define 
\begin{equation*}
\footnotesize
    \begin{aligned}
    W=\sqrt{ (1-\frac{\beta}{g-\beta})(1+\frac{\beta}{g-\beta}) } \cdot \sqrt{(1-\frac{\beta}{m})(1+\frac{\beta}{m-2\beta})},
    \end{aligned}
\end{equation*}
then
\begin{equation*}
\footnotesize
    \begin{aligned}
    B_n&=W\cdot (1-\frac{\beta}{g-\beta})^{g-\beta}(1+\frac{\beta}{g-\beta})^{g-\beta}\\
    &\cdot (1-\frac{2 \beta}{g})^{\beta}(1-\frac{\beta}{m-\beta})^{-(m-\beta)}(1+\frac{\beta}{m-\beta})^{-(m-\beta)} (1-\frac{2\beta}{m})^{\beta}\\
    &=W\cdot e^{-\beta}\cdot e^\beta\cdot e^{\frac{2\beta^2}{g}}\cdot e^{\beta}\cdot e^{-\beta}\cdot e^{-\frac{2\beta^2}{m}}\\
    &= W\cdot  \exp\{ \frac{2\beta^2(m-g)}{g\cdot m} \}.
    \end{aligned}
\end{equation*}

Substituting $g=\frac{n(n-1)}{2}\approx \frac{n^2}{2}$ and $m=g\cdot n^{-\frac{2c}{k-1}}$, we derive:

\begin{equation*}
\footnotesize
\begin{aligned}
\frac{2\beta^2(m-g)}{g\cdot m}&=\frac{2\beta^2(g\cdot n^{-\frac{2c}{k-1}}-g)}{g\cdot g\cdot n^{-\frac{2c}{k-1}}} =\frac{2\beta^2(1-n^{\frac{2c}{k-1}})}{g}.
\end{aligned}
\end{equation*}

Under the condition $c<1$, we have $\frac{2c}{k-1}<2$. Consequently, $n^{\frac{2c}{k-1}}=o(n^2)$. This implies:
\begin{equation*}
\footnotesize
\left|\frac{2\beta^2(1-n^{\frac{2c}{k-1}})}{g}\right| \leq \frac{2\beta^2n^{\frac{2c}{k-1}}}{\frac{n^2}{2}}=\frac{4\beta}{n^{2-\frac{2c}{k-1}}} \to 0.
\end{equation*}

Thus, $\exp\{ \frac{2\beta^2(m-g)}{g\cdot m} \}=1$. Additionally, it is straightforward to verify $W=1 + o(1)$. Therefore, we conclude $\lim\limits_{n \to \infty} B_n=1$. 

For (\ref{C_n}), we have

\begin{equation*}
\footnotesize
\begin{aligned}
C_n &= \binom{n}{k}^{-1} \sum_{r=2}^k \binom{k}{r} \binom{n - k}{k - r} \binom{g-w}{m-w} \binom{g}{m} \binom{g-\beta}{m-\beta}^{-2},
\end{aligned}
\end{equation*}

following the proof method mentioned above, when $0 < k < \ln(n), r<k, n \to \infty$ it can be readily shown that:

\begin{equation*}
\footnotesize
    \begin{aligned}
    C_n &= \sum_{r=2}^k \binom{k}{r} \binom{g-w}{m-w} \binom{g}{m} \binom{g-\beta}{m-\beta}^{-2}\\
    &\leq \sum_{r=2}^k(\frac{ek^2}{rn})^r \binom{g-w}{m-w} \binom{g}{m} \binom{g-\beta}{m-\beta}^{-2}\\
    &= \sum_{r=2}^k(\frac{ek^2}{rn})^r (\frac{g-w}{g-\beta})^{g-\beta} (\frac{g}{g-\beta})^{g-\beta} \\
    &\cdot g^{\beta}(g-w)^{\beta-w}(\frac{m}{m-\beta})^{-(m-\beta)}\\
    &\cdot (\frac{m-w}{m-\beta})^{-(m-\beta)}(\frac{1}{m})^\beta(\frac{1}{m-w})^{\beta-w} \\
    &= \sum_{r=2}^k(\frac{ek^2}{rn})^r e^{\beta-w}e^{\beta}e^{-\beta}e^{w-\beta}(\frac{g}{m})^{\beta}(\frac{g-w}{m-w})^{\beta-w}\\
    &= \sum_{r=2}^k(\frac{ek^2}{rn})^r (\frac{g}{m})^{\beta}(\frac{g-w}{m-w})^{\beta-w}.
    \end{aligned}
\end{equation*}

Assume the individual term attains its maximum at $r=u$. Then:
\begin{equation*}
\footnotesize
    \begin{aligned}
    C_n &\leq k \left( \frac{ek^2}{un} \cdot  (n^{\frac{2c}{k-1}})^{\frac{u-1}{2}} \right)^u\\
    &= k\left( \frac{ek^2}{u}\cdot n^{\frac{c(u-1)-(k-1)}{k-1}}\right) ^u\\
    &\leq k(ekn^{\frac{(c-1)(k-1)}{k-1}})^k\\
    &=e^kk^{k+1}n^{k(c-1)}.
    \end{aligned}
\end{equation*}

Substituting $m=g\cdot n^{-\frac{2c}{k-1}}$, then, 
\begin{equation*}
\footnotesize
    \begin{aligned}
    C_n &\leq k \left( \frac{ek^2}{un} \cdot  (n^{\frac{2c}{k-1}})^{\frac{u-1}{2}} \right)^u\\
    &= k\left( \frac{ek^2}{u}\cdot n^{\frac{c(u-1)-(k-1)}{k-1}} \right) ^u\\
    &\leq k(ekn^{\frac{(c-1)(k-1)}{k-1}})^k\\
    &=e^kk^{k+1}n^{k(c-1)}.
    \end{aligned}
\end{equation*}

For $c < 1$, we have $\lim\limits_{n \to \infty} C_n = 0$. Thus, the equation (\ref{E_x_2_equ}) can be obtained.

When $m > \frac{n(n-1)}{2} \cdot n  ^{-\frac{2}{k-1}}$, by (\ref{E_x_equ}) for $\mathbb{E}[X] \to \infty$

\begin{equation*}
\footnotesize
\label{E_x_2_var}
    \begin{aligned}
    \frac{Var(X)}{\mathbb{E}^2[X]}
    = \frac{1}{\mathbb{E}[X]} + A_n \cdot  B_n + C_n - 1 = \frac{1}{\mathbb{E}[X]}.
    \end{aligned}
\end{equation*}

This implies
\begin{equation*}
\footnotesize
  \lim\limits_{n \to \infty}Var(X) = o(\mathbb{E}^2[X]),
\end{equation*}
which rigorously establishes that $k$-clique exist with high probability in this regime.

\end{proof}

Therefore, we conclude that
\begin{equation}
\footnotesize
m=\frac{n(n-1)}{2}\cdot n^{-\frac{2}{k-1}}
\end{equation}
marks the phase transition phenomenon for $k$-clique emergence. 

\textbf{Notation.}
Following the same strategy as \cite{xu2025sat}, we can pick asymptotic value $m=\frac{n(n-1)}{2}\cdot n^{-\frac{2}{k-1}}+\epsilon_n$ (where $\epsilon_n$ tends to 0) such that $\mathbb{E}[X]=1/2$, then $\mathbb{P}(X>0)\ge\mathbb{E}[X]^2/\mathbb{E}{[X^2]}=\frac{1}{\frac{Var(X)}{\mathbb{E}^2[X]}+1} \approx\frac{1}{1/\mathbb{E}[X]+1} \ge \frac{1}{3}$. Therefore, $\mathbb{P}(X=1)\ge1/6$.

\section{Self-referential instances}

Locating a $k$-clique or determining its non-existence in the phase transition point typically requires substantial computational time. This phenomenon has motivated our investigation into hard instances. Since many existing algorithms employ heuristics based on vertex degrees, this paper constructs challenging instances by focusing on degree properties. 

\begin{definition}
A symmetric transformation refers to an operation that alters the adjacency relationships in a graph while preserving the degree of every vertex. 
\end{definition}

\begin{figure}[htp]
 \centering
 \includegraphics[width=8cm]{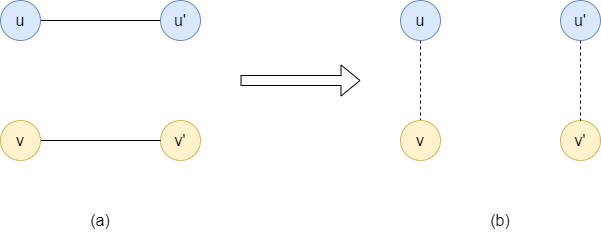}
 \caption{In Figure (a), the solid lines represent the original edges in the graph, and each vertex has a degree of 1. After the transformation, the new graph is shown in Figure (b). Following the establishment of these new edges, each vertex still maintains a degree of 1, thereby ensuring degree invariance. }
 \label{fig:fig_turn_A}
\end{figure}

This transformation process is illustrated in Figure \ref{fig:fig_turn_A}. After this transformation, the degrees of all vertices remain unchanged, and the newly generated edges did not exist in the original graph. Using this method, we can disrupt the clique structure. Similarly, this process is reversible—we can also establish clique structures through the reverse symmetric transformation. 

Thus, we can derive the following theorem. 

\begin{theorem}
There exists a set of graphs in which every graph shares an identical number of vertices, number of edges, and vertex degree sequence. This set includes both instances that contain a $k$-clique and instances that do not contain any $k$-clique. Moreover, the set maintains self-referential nature under the symmetric transformation, meaning this transformation enables mutual conversion between instances with and without a solution.
\end{theorem}

\begin{proof}
Suppose $G=(V, E)$ contain exactly one $k$-clique $S=(V_S, E_S)$ (such $S$ was shown to exist, as established in the notation introduced at the end of Section 3). Assume the original clique $S$ has size $k$. Suppose there exists an edge $(u, u') \in E_S$ and another edge $(v, v') \notin E_S$ but $(v, v') \in E$, where $(u, v) \notin E$ and $(u', v') \notin E$.

Construct a modified graph $G'=(V, E')$ by removing edges $(u, u')$ and $(v, v')$ and adding edges $(u, v)$ and $(u', v')$. This operation preserves both the total number of edges and the degree sequence of $G$. 

After deleting $(u, u')$, $S$ splits into two disjoint subsets $S_1$ and $S_2$, each forming a $(k-1)-clique$ if the remaining adjacency is preserved. 

If $\{ u, v, w_1, . . . , w_{k-2} \}$ forms a new $k$-clique $G'$, then it must satisfy that$(v, w_i) \in E$ for $\forall i \in \{1, . . . , k-2\}$. Similarly, if
$\{u', v', w_1, . . . , w_{k-2}\}$ forms a new $k$-clique, then $(v', w_i) \in E$ for $\forall i \in \{1, . . . , k-2\}$. 

The probability that the modified graph $G'=(V', E')$ still contains a $k$-clique is given by:

\begin{equation*}
\footnotesize
 \begin{aligned}
 P &= \frac{2 \binom{g-\beta-(k-2)}{m-\beta-(k-2)} - \binom{g-\beta-2(k-2)}{m-\beta-2(k-2)} }{\binom{g-\beta}{m-\beta}}\\
 &=2 \sqrt{1+\frac{g-m}{m-\beta-k+2}}(\frac{g-\beta-k+2}{g-\beta})^{g-\beta} \cdot (\frac{1}{g-\beta-k+2})^{k-2}(\frac{m-\beta}{m-\beta-k+2})^{m-\beta}\\
 &\cdot (m-\beta-k+2)^{k-2}-\sqrt{1+\frac{g-m}{m-\beta-2k+4}}
 \cdot (\frac{g-\beta-2k+4}{g-\beta})^{g-\beta}(\frac{1}{g-\beta-2k+4})^{2k-4}\\
 &\cdot (\frac{m-\beta}{m-\beta-2\beta+4})^{m-\beta}(m-\beta-2k+4)^{2k-4}\\
 &= 2 \sqrt{1+\frac{g-m}{m-\beta-k+2}}(1+\frac{m-g}{g-\beta-k+2})^{k-2}- \sqrt{1+\frac{g-m}{m-\beta-2k+4}}(1+\frac{m-g}{g-\beta-2k+4})^{2k-4}.\\
 \end{aligned}
\end{equation*}

Note that

\begin{equation*}
\footnotesize
\begin{aligned}
A \triangleq \sqrt{1+\frac{g-m}{m-\beta-k+2}}(1+\frac{m-g}{g-\beta-k+2})^{k-2}=o(1)\\
B \triangleq \sqrt{1+\frac{g-m}{m-\beta-2k+4}}(1+\frac{m-g}{g-\beta-2k+4})^{2k-4}=o(1).
\end{aligned}
\end{equation*}



Then
\begin{equation*}
\footnotesize
\begin{aligned}
\lim\limits_{n \to \infty} P&= 0.
\end{aligned}
\end{equation*}

Therefore, after symmetric transformation, the resulting new graph $G'$ will not contain any $k$-cliques with high probability. This implies the existence of a particular vertex selection and corresponding edge changes that transform a solvable instance (containing a $k$-clique) into an unsolvable one.

Similarly, for graphs initially without a $k$-clique, we may assume there exists a $(k-1)$-clique $S$. (The existence of such an $S$ follows from standard second-moment arguments on the number of $(k-1)$-sets that cliques. We omit the routine but lengthy details here). Then select a vertex $v \notin S$ and a vertex $u' \in S$ such that $(u, u') \notin E$. 

Choose edges $(u, v') \in E , (u', v) \in E, (v, v')\notin E$. Modify $G'$ by removing $(u, v')$ and $(u', v)$, then adding $(u, u')$ and $(v, v')$ to form a new graph $G'=(V, E')$. 

For a new $k$-clique to emerge, the remaining $k-2$ vertices in $S/{u'}$ must be adjacent to $u$ or $v$. 

The probability of this event is also bounded by $P$. Since $P = o(1)$, there exists a vertex selection that can transform an unsolvable instance into a solvable one. 

\end{proof}

\begin{figure}[htp]
 \centering
 \includegraphics[width=8cm]{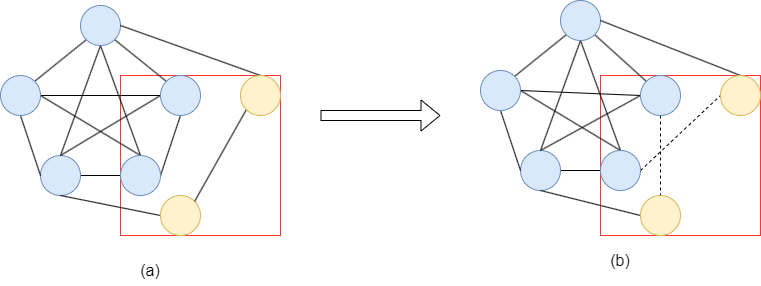}
 \caption{In figure (a), the blue vertices form a clique of size 5. After the dashed edges are added in figure (b), the clique formed by the blue vertices is disrupted and cannot form a new clique. }
 \label{turn_A_2}
\end{figure}

As shown in Figure \ref{turn_A_2}, we perform transformations on the vertex pairs within the red box. After removing the solid edges and establishing new dashed edges, the degrees of the four affected vertices remain unchanged, while the original clique structure is disrupted. Therefore, this transformation method can be used to construct two types of graphs: one containing a $k$-clique and another without a $k$-clique. 

Such transformations can disrupt or establish clique structures without altering the graph's characteristics. Regarding current clique-finding algorithms—taking Branch and Bound as an example—most are based on neighborhood search algorithms, meaning vertex degrees serve as a crucial evaluation factor for the initial search starting point. Therefore, this paper, considering the perspective of vertex degrees, utilizes the self-referential property of graphs to perform transformations that disrupt or establish cliques within the graph.

If we represent the search space as a set of graphs with the same number of vertices, the same number of edges, and the same degree for each vertex, then the algorithm’s search process continually reduces this set. If the set eventually contains only one state—either solvable or unsolvable—the search can terminate early and return the result. However, if the set still contains both solvable and unsolvable possibilities, the algorithm must continue searching. Next, we will use an example to illustrate why all candidate solutions must be nearly almost examined to determine whether a solution exists.

\begin{figure}[htp]
 \centering
 \includegraphics[width=8cm]{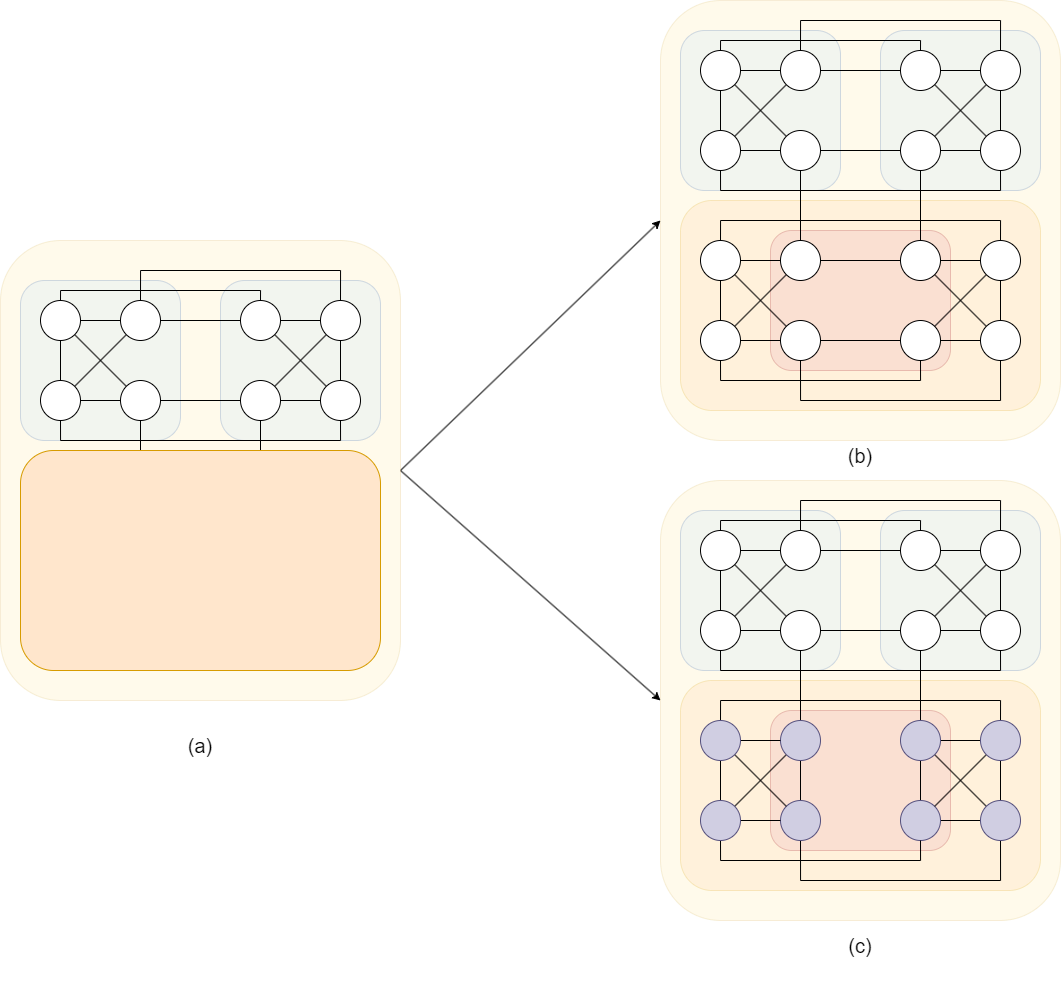}
 \caption{Figure (a) illustrates the current search state of the program. The blue region represents the portion of the graph that has already been searched, while the orange region indicates areas not yet explored. Within the possible set of graphs, two distinct configurations exist: Figure (b) and Figure (c). Figure (b) does not contain a 4-clique, whereas Figure (c) does (formed by the purple vertices). Figures (b) and (c) can be obtained from each other by applying a symmetric transformation within the red region.}
 \label{no_ans}
\end{figure}

In a local section of a graph, we may identify numerous candidate solutions. However, until the search over these candidates is exhaustive, we cannot definitively determine whether the entire graph is solvable or unsolvable. As illustrated in Figure \ref{no_ans}, the blue area represents the region that has been searched, while the orange area denotes the unexplored portion. Within the unexplored region, a symmetric transformation can potentially convert an unsolvable configuration into a solvable one. Consequently, two distinct possibilities exist within the graph set, necessitating further search by the algorithm to ascertain whether the graph contains a $k$-clique.

When performing a symmetric transformation in the phase transition region, it is possible to avoid affecting the already searched space. The fundamental reason lies in the independence between candidate solutions. Adjusting some candidate solutions does not impact the structure of other candidate solutions.

When performing multiple symmetric transformations, we observe that this transformations operation does not affect the structure of the already searched portions. We attribute this phenomenon to the independence between candidate solutions. Specifically, consider two candidate solutions $\sigma$ and $\tau$, with the number of overlapping vertices being $r$. This scenario aligns with the situation described in Lemma \ref{Geq_phase}, where $A_n$ denotes the probability of overlap size $0$ or $1$, and $C_n$ denotes the probability of overlap size $2$. Subsequent calculations show that when $r>1$, the probability of such an event almost is zero.

We have
$$
P(\sigma,\tau\text{ are solutions}) \approx P(\sigma\text{ is a solution}) \cdot P(\tau\text{ is a solution}).
$$
That is, the probability that both candidate solutions are valid solutions simultaneously and the product of the probabilities that each candidate solution is a solution separately tend to become equal as $n$ grows. This indicates that the candidate solutions are almost mutually independent, which also explains why symmetric transformations do not affect the structure of other candidate solutions.

\section{Conclusions}

In this paper, we construct self-referential instances for the well-known clique problem, following a similar approach to that proposed by Xu and Zhou~\cite{xu2025sat}. We prove the existence of a phase transition phenomenon in the clique problem and compute the location of the phase transition point. Subsequently, we investigate the hardness phenomenon in the phase transition point. By leveraging the self-referential property of the clique problem, we demonstrate that in the phase transition point, one can transform the satisfiability of an instance without altering the graph’s characteristics. This shows that at the phase transition point, there exists a class of instances for which confirming whether the graph has a solution requires exhaustive search. This observation is consistent with previous results on the maximum independent set problem within the framework of parameterized complexity~\cite{chen2006strong}. Furthermore, from the perspective of independence within the solution space, we clarify two key points: why self-referential instances can be constructed, and why exhaustive search becomes indispensable for them—both of which are rooted in the fact that candidate solutions are almost mutually independent. This is also the intrinsic reason why Xu and Zhou~\cite{xu2025sat} can use the RB model to construct self-referential instances. We expect this approach to be applied to more relevant problems, explaining why these problems require exhaustive search.

\nocite{*}

\printbibliography
\end{CJK}
\end{document}